\documentclass[preprint,showpacs,amsmath,amssymb]{revtex4}
\usepackage{graphicx}
\usepackage{dcolumn}
\usepackage{bm}
\usepackage{color}

\usepackage{here}

\begin{document}

\title{Numerical study on a canonized Hamiltonian system representing reduced magnetohydrodynamics and its comparison with two-dimensional Euler system}

\author{Yuta Kaneko and Zensho Yoshida}

\affiliation{%
         Graduate School of Frontier Sciences, The University of Tokyo, Kashiwa, Chiba 277-8561, Japan}

\begin{abstract}
Introducing a Clebsch-like parameterization, we have formulated a canonical Hamiltonian system on a symplectic leaf of reduced magnetohydrodynamics.
An interesting structure of the equations is in that the Lorentz-force, which is a quadratic nonlinear term in the conventional formulation, appears as a linear term $-\Delta Q$, just representing the current density ($Q$ is a Clebsch variable, and $\Delta$ is the two-dimensional Laplacian);
omitting this term reduces the system into the two-dimensional Euler vorticity equation of a neutral fluid.
A heuristic estimate shows that current sheets grow exponentially (even in a fully nonlinear regime) 
together with the action variable $P$ that is conjugate to $Q$.
By numerical simulation, the predicted behavior of the canonical variables, yielding exponential growth of current sheets, has been demonstrated.
\end{abstract}

\pacs{52.35.We,45.20.Jj}

\maketitle  

\section{Introduction}

The reduced magnetohydrodynamics (RMHD) \cite{refBB, refMN, refHR76, refHR77} is a powerful model to describe macroscopic nonlinear phenomena of a plasma on a two-dimensional cross section perpendicular to a strong, almost homogeneous, longitudinal magnetic field.
The system of RMHD equations is an extension of the two-dimensional Euler vorticity equation (2DEV);
a quadratic nonlinear term of magnetic flux ($\psi$), representing the magnetic force, is added to 2DEV,
while the evolution of $\psi$ is governed by the flow to be determined by the vorticity equation. 

Both 2DEV and RMHD systems can be cast into Hamiltonian formalisms\,\cite{refPJ80, refPJ84, refM},
which, however, are \emph{noncanonical} in the sense that the Poisson operators have nontrivial kernels.
Hence, the phase space of the Eulerian variables is \emph{foliated}.
The Clebsch parametrization\,\cite{refAC,refZY09} is an effective method to canonize the system\,\cite{refM,refZY12,refYH2013}. 

In our previous study~\cite{refYK}, we derived a variety of canonized systems of RMHD by extending the Clebsch parameterization.
An interesting observation was that the nonlinear magnetic force term is represented differently in terms of the canonical variables parameterizing of the vorticity $\omega$ and the magnetic flux $\psi$;
in the simplest form, it becomes a linear term.
In this paper, we invoke this simplest system to elucidate the role of magnetic-force term
both analytically and numerically.
The creation of ``current sheet'' is the main subject of study;
we can measure the current by $-\Delta\psi$
($\Delta$ is the two-dimensional Laplacian), where 
$\psi$ is one of the Clebsch parameters, and $-\Delta\psi$ is directly the magnetic-force term in the
canonized formulation. 
If we omit the magnetic-force term, the system degenerates into 2DEV.
In RMHD, $\psi$ is a physical field (i.e. the magnetic flux),
and the magnetic force $-\Delta\psi$ influences the dynamics.
In 2DEV, however, $\psi$ is an abstract Clebsch parameter, which 
is involved in the dynamics only through its contribution in the parameterization of $\omega$.
We observe that the magnitude of $-\Delta\psi$ becomes stronger in RMHD in comparison with 2DEV.

In Sec.\,\ref{sec:formulation}, we start by formulating a canonized Hamiltonian system of RMHD.
In Sec.\,\ref{sec:scaling}, we make a heuristic estimate of the growth rate of a current sheet in the canonized system. 
In Sec.\,\ref{sec:experiment}, we describe the result of numerical simulations, 
and examine the prediction of Sec.\,\ref{sec:scaling}. 
Section\,\ref{sec:conclusion} concludes the study.

\section{Canonized Hamiltonian dynamics of Reduced Magnetohydrodynamics}
\label{sec:formulation}
On a two-dimensional flat plane, we consider a compact periodic domain.
We define $[a,b]=-\nabla a \times \nabla b \cdot\bm{e}_z$,
where $\bm{e}_z$ is the unit normal vector onto the plane.
We consider a coupled nonlinear equations
\begin{align}
&\partial_t \omega + [\phi, \omega ]= \alpha [\psi,J], \label{624-2} \\
&\partial_t \psi + [\phi, \psi]=0, \label{624-1}
\end{align}
where $\phi$ is the Gauss potential of the flow, $\omega=-\Delta\phi$ is the vorticity,
$\psi$ is the magnetic flux, $J=-\Delta \psi$ is the current density,
and $\alpha$ is a constant measuring the magnetic field strength.
Putting $\alpha=1$ yields the RMHD equations in the standard Alfv\'en units.
On the other hand, putting $\alpha=0$ eliminates the magnetic-force term, 
and then, (\ref{624-2}) reduces into 2DEV;
(\ref{624-1}) is decoupled, and $\psi$ becomes a passive scalar.

The energy of the system is 
\[
H= \frac{1}{2}\int (V^2 + \alpha B^2)\,\mathrm{d}^2 x,
\]
where $\bm{V}=\nabla\phi\times\bm{e}_z$ is the flow velocity, and $\bm{B}=\nabla\psi\times\bm{e}_z$ is
the (poloidal) magnetic field.
Representing $H$ in terms of $\omega$ and $\psi$,
we define the Hamiltonian as
\begin{equation}
H(\omega,\psi) = -\frac{1}{2} \int \left( \Delta^{-1} \omega \cdot \omega + \alpha \Delta \psi \cdot \psi \right) \mathrm{d}^2 x.
\label{Hamiltonian}
\end{equation}
Here, $\Delta^{-1}$ is the inverse operator of $\Delta$, i.e. $\Delta^{-1}\omega=\phi$
with setting the gauge of the Gauss potential $\phi$ to be zero ($\int\phi \mathrm{d}^2x=0$),
which is a self-adjoint operator.
Henceforth, the phase space is the $L^2$ Hilbert space of $\bm{u}= ~^t(\omega,\psi)$.
Introducing a Poisson operator
\begin{align}
\mathcal{J}(\omega,\psi) =
  \left(
    \begin{array}{cc}
          \left[ \omega, \circ \right] & \left[ \psi, \circ \right] \\
          \left[ \psi,\circ \right] & 0 
    \end{array}
  \right)
  \label{Poission-operator}
\end{align}
($\circ$ means insertion), we can cast the RMHD equations (\ref{624-2}) and (\ref{624-1})
into a Hamiltonian form
\begin{align}
  \partial_t \left(
    \begin{array}{c}
           \omega \\
           \psi
    \end{array}
  \right)
  &= \mathcal{J}(\omega,\psi) \left(
    \begin{array}{c}
          \partial_\omega H \\
          \partial_{\psi} H
    \end{array}
  \right). \label{1030-1}
\end{align}
The corresponding Lie-Poisson bracket\,\cite{refPJ80, refPJ84, refM}
\begin{align}
  \{ F,G \} &= \langle \partial_{\bm{u}} F, \mathcal{J} \partial_{\bm{u}} G\rangle
  \nonumber \\
            &= \int  \: W_{ij} \left[ \partial_{u_i}F, \partial_{u_j}G \right] \mathrm{d}^2x, 
  \label{624-4}
\\
W_{ij} &= -\left(
    \begin{array}{cc}
           \omega & \psi \\
        \psi & 0
    \end{array}
  \right)
\end{align}
defines a ``degenerate'' Poisson algebra on the phase space;
the Casimir elements (a functional $C(\bm{u})$ satisfying $\{C, G\}=0$ for every $G(\bm{u})$ is
called a Casimir element, which is, therefore, a constant of motion) are  
\begin{align}
C_1(\bm{u})= \int f(\psi) \mathrm{d}^2 x ,
\quad
C_2(\bm{u})= \int \omega g(\psi) \mathrm{d}^2 x, \label{8}
\end{align}
where $f$ and $g$ are arbitrary functions~\cite{refPJ84, refPe2}.
The constancy of $C_1$ implies the flux conservation for every magnetic surfaces.
The functional $C_2$ (especially with $g(\psi)=\psi$) is called the \emph{cross helicity},
which can be related to a N\"other charge pertinent to a redundancy of the Lagrangian formalism\,\cite{refM,refYF}.

By an appropriate parameterization of the state vector,
we can derive a canonized subsystem on a submanifold of the Poisson manifold.
In \cite{refYK}, a variety of canonized systems were formulated.  
Here we invoke a simple one:
\begin{align}
& \omega = [P,Q] ,
 \label{c01-1} \\ 
& \quad \psi =  Q .
 \label{c01-2} 
\end{align}
Notice that $\omega$ is written as a Clebsch 2-form~\cite{refZY09}.
Evidently, 
by the periodic boundary condition,
\begin{align}
C_2= \int g(Q)[P,Q] \mathrm{d}^2x=
\frac{1}{2}\int [P,G(Q)] \mathrm{d}^2x=0,
\label{cross-helicity-QP-form}
\end{align}
where $G(Q)'=g(Q)$.
Hence, the manifold spanned by $Q$ and $P$ is a zero cross helicity Casimir leaf;
on which we can introduce a symplectic structure.
We consider a canonical Hamiltonian system
\begin{align}
  \partial_t \left(
    \begin{array}{c}
           Q \\
           P
    \end{array}
  \right)
  = \mathcal{J}_C \left(
    \begin{array}{c}
          \partial_Q H \\
          \partial_{P} H
    \end{array}
  \right) ,
\label{703-1}
\end{align}
where $\mathcal{J}_C$ is a symplectic operator:
\begin{align}
\mathcal{J}_C =  \left(
    \begin{array}{cc}
         0 & I \\
        -I  & 0
    \end{array}
    \right).
\end{align}
The Hamiltonian is now written in terms of the canonical variables:
\begin{align}
H(Q,P) = -\frac{1}{2} \int \left( \Delta^{-1} [P,Q] \cdot [P,Q] + \alpha \Delta Q \cdot Q \right) \mathrm{d}^2 x.
\label{Hamiltonian2}
\end{align}
Explicitly, we may write (\ref{703-1}) as
\begin{align}
& \partial_t Q = -\left[ \phi, Q \right] ,
\label{703-1-Q}
\\
& \partial_t P = -\left[ \phi,P  \right] + {\alpha} \Delta Q .
\label{703-1-P}
\end{align}
Here, we have evaluated $\Delta^{-1}[P,Q]=\phi$ with setting the gauge of the
Gauss potential $\phi$ to be zero.  

We can easily verify that $\omega =[P,Q]$ and $\psi=Q$, in terms of the canonical variables
$Q$ and $P$ obeying (\ref{703-1}),
satisfy the original noncanonical Hamiltonian system (\ref{1030-1}),
i.e. (\ref{703-1}) describes the canonical Hamiltonian mechanics on the $Q$-$P$ symplectic leaf.
In comparison with the original system (\ref{624-2})-(\ref{624-1}),
it is remarkable that the magnetic-force term (measured by $\alpha$) 
is ``linearized'' in (\ref{703-1-Q})-(\ref{703-1-P}).

\section{A scaling of current sheet}
\label{sec:scaling}
We derive a heuristic estimate of the growth rate of a local current density $J=-\Delta Q$ (here we put $\alpha=1$). 
Suppose that a \emph{current sheet} is created; 
by a current sheet, we mean a narrow region, supporting peaked $J$, which stays almost stationary at a fixed place.  

Since $Q$ obeys the \emph{Liouville equation} (\ref{703-1-Q}), the maximum value of $|Q|$ does not change.
The growth of $J=-\Delta Q$ is due to the reduction of the thickness of the current sheet caused by the convection.
Let $x$ be the local coordinate in the direction perpendicular to the sheet; we denote by $y$ the parallel coordinate.
A typical incompressible flow that can cause such convection is given by, in the vicinity of the current sheet,
\begin{equation}
\phi = \phi_0 = -\nu(t) x y,
\label{phi_0}
\end{equation}
which yields a velocity field $V_x=-\nu(t) x$, $V_y=\nu(t) y$
($\nu(t)$ is a certain real-valued function of $t$).  
Notice that this flow is irrotational ($\nabla\times\bm{V}=0$).
Solving the characteristic equation $d\bm{x}/dt = \bm{V}$, we obtain the streamlines:
denoting the Lagrangian coordinates $x(0)=x_0$ and $y(0)=y_0$,
\begin{equation}
x(t) = e^{-\int_0^t \nu(\tau)d\tau}x_0, \quad y(t) = e^{\int_0^t \nu(\tau)d\tau}y_0 .
\label{characteristics}
\end{equation}
By the pull-back map $(x,y,t) \mapsto (x_0,y_0,0)$, 
the d'Alembert solution to (\ref{703-1-Q}) is given by
\begin{eqnarray}
Q(x,y,t) &=& Q_0 (x_0(x,y,t), y_0(x,y,t)) 
\nonumber \\
&=& Q_0(e^{\int_0^t \nu(\tau)d\tau}x,e^{\int_0^t \nu(\tau)d\tau}y),
\label{dAlembert-0}
\end{eqnarray}
where $Q_0$ is the initial distribution of $Q$.

We begin with $\nu(t)=v_0$, a positive constant.
Then, an exponential scale change occurs:
\[
x(t)=e^{-v_0 t}x_0, \quad y(t)=e^{v_0 t}y_0.
\]
The d'Alembert solution (\ref{dAlembert-0}) reads
\begin{equation}
Q(x,y,t) = Q_0(e^{v_0 t}x,e^{-v_0 t}y),
\label{dAlembert}
\end{equation}
for which we may estimate
\begin{equation}
\partial_x \sim e^{v_0 t}, \quad \partial_y \sim e^{-v_0 t},
\label{scating_linear}
\end{equation}
thus
\begin{equation}
J = -\Delta Q \approx e^{2v_0 t} J_0,
\label{scating_linear_Q}
\end{equation}
where $J_0=-\Delta Q_0$.

Let us examine the consistency of the estimate (\ref{scating_linear_Q}) with the evolution equation of $J$.
Differentiating (\ref{703-1-Q}), we obtain
(denoting $\partial_x f = f_x$ and $\partial_y f = f_y$)
\begin{equation}
\partial_t J = [\phi,J ] -  [\Delta\phi, Q] 
- 2 [\phi_x,Q_x] - 2 [\phi_y,Q_y ].
\label{J-equation}
\end{equation}
In the current sheet, we may approximate $J \approx - Q_{xx}$.
Since we assume that the current sheet does not move around,
the convection term $[\phi,J ]$ vanishes there.
Assuming (\ref{phi_0}), we may put $\Delta\phi=0$.
On the right-hand side of (\ref{J-equation}), then, only the term $2[\phi_x,Q_x] = 2\phi_{xy}Q_{xx}$ 
must be retained; (\ref{J-equation}) now reads
\begin{equation}
\partial_t J \approx -2\phi_{xy} J = 2 v_0 J.
\label{J-equation-approx}
\end{equation}
Integrating (\ref{J-equation-approx}) yields (\ref{scating_linear_Q}).

The foregoing estimate is based on an \emph{a priori} Gauss potential $\phi=\phi_0$.
We have yet to adjust $\phi$ to make it consistent with $P$ and $Q$.
Let us put $\phi = \phi_0 + \tilde{\phi}$.
By the definition (\ref{c01-1}), we have
\[
 -\Delta \phi = -\Delta \tilde{\phi} = [P,Q].
\]
To determine $P$ in the current sheet, we integrate (\ref{703-1-P});
there, the convection term $[\phi,P]$ is negligible.
For the $J=-\Delta Q$ of (\ref{scating_linear_Q}), we obtain
\begin{align}
& P \approx P_0 + \int_0^t J ~dt \approx P_0 + \frac{e^{2v_0 t}}{2v_0} J_0.
\label{P-estimate}
\end{align}
Now, the determining equation (\ref{J-equation}) of $J$ includes an additional leading-order term pertinent to $\Delta\phi=[P,Q]$, i.e. 
we have to modify (\ref{J-equation-approx}) as
(neglecting $P_0$ with respect to the exponentially growing term $(2v_0)^{-1}e^{2v_0 t} J_0$)
\begin{align}
\partial_t J & \approx -2\phi_{xy} J + [[P,Q],Q] 
\nonumber \\
& \approx 2 v_0 J + \frac{e^{2v_0t}}{2v_0}[[J_0,Q],Q].
\label{J-equation-approx-2}
\end{align}
By the scaling (\ref{scating_linear}), we may assume that $[[J_0,Q],Q]$ remains constant in the current sheet
(notice that the bracket $[~,~]$ is a bilinear form always pairing $\partial_x$ and $\partial_y$).
Integrating (\ref{J-equation-approx-2}) yields
\begin{equation}
J \approx ( 1 + \varepsilon t) e^{2v_0 t} J_0 ,
\quad \varepsilon = \frac{[[J_0,Q],Q]}{2v_0 J_0}.
\label{scating_secular_J}
\end{equation}
Iterating this $J$ into (\ref{P-estimate}), 
the secular term ($\varepsilon t$) yields a renormalized time constant:
\begin{equation}
J \approx e^{2v'_0 t} J_0 ,
\quad v'_0 = v_0 + \frac{\varepsilon}{2}.
\label{scating_renormalized_J}
\end{equation}
Notice that $\varepsilon=0$, if we put $\alpha=0$ in (\ref{703-1-P}).

We note that $\varepsilon$ is not necessarily a small number.
In fact, the numerical simulation, described in the next section, shows that $\varepsilon$
is large, causing a stronger peaking of $J=-\Delta Q$ in RMHD ($\alpha=1$) compared with 2DEV ($\alpha=0$).

We end this section with a remark on the ``exponential growth'' of the current.
While such behavior in RMHD systems is commonly observed in simulations studies~\cite{refDC,refRG},
it might be thought peculiar that the evolution of the nonlinear system is exponential.
As shown in the d'Alembert solution (\ref{dAlembert-0}), the growth of $J=-\Delta Q$ is
caused by the scale reduction due to the flow $\bm{V}=\nabla\phi\times\bm{e}_z$.  
If the Gauss potential $\phi$ was independent of $Q$, the determining equation (\ref{703-1-Q}) of $Q$
is a linear evolution equation, thus the growth of $J=-\Delta Q$ is naturally exponential;
the first part of the foregoing analysis delineates this fact.
In the latter part, we have studied the self-consistent relation between $\phi$ and the dynamical variables $Q$ and $P$.
The key element of construction was the integration (inside the current sheet) 
of the determining equation (\ref{703-1-P}) of $P$,
which includes $J$ as an inhomogeneous term.
For $J\propto e^{\gamma t}$ ($\gamma=2v'_0$), $P\approx \int J dt \propto e^{\gamma t}$, and then,
the determining equation (\ref{J-equation-approx-2}) of $J$ yields the consistent $J\propto e^{\gamma t}$.
At the core of the self-consistency is the fact that the integral of an exponential function is a similar exponential function.
The integral $\int J dt$ serves as a secular perturbation accelerating the growth of $J$ itself through the term including $P$.
If $J$ is an exponential function of $t$, the whole terms of the evolution equation (\ref{J-equation-approx-2}) can balance.
A different type of evolution cannot maintain the consistency.
Intuitively we may understand that integration boosts a slow growth (given by, for example, a polynomial), but suppresses a rapid change;
the exponential growth is just at the balance.
For instance, if we assume a faster growth such as $J\propto \exp [a e^{b t}]$,
we observe
\begin{align}
\int_0^t \exp [a e^{b \tau}]d\tau &= b^{-1} \textrm{Ei}(a e^{bt})
\nonumber \\
&\rightarrow b^{-1} \exp [ a e^{bt} -bt] \quad (t\rightarrow\infty) .
\nonumber
\end{align}
Hence, the growth of $P$ becomes slower by factor $e^{-bt}$ than $J$, destroying the similarity
of the fields constituting the current sheet.

\section{Numerical experiment}
\label{sec:experiment}
By numerical simulation, we study the nonlinear evolution of the canonized RMHD system ($\alpha=1$), 
and compare it with the 2DEV system ($\alpha=0$).
We consider a $2\pi \times 2\pi$ square domain with the periodic boundary condition.
We invoke the pseudo spectral method to deal with the spatial derivatives and nonlinear terms.
The temporal derivatives are discretized by the fourth-order Runge-Kutta method.

We assume an initial condition
\begin{equation}
\left\{ \begin{array}{l}
Q(x,y,0)=\cos (x+1.7) - \sin (y+3.7) ,
\\
P(x,y,0)=\sin (x+1.5) \sin (y+2.3)+ \cos (y+4.7).
\end{array} \right.
\label{IC}
\end{equation}
Here, the variables are written in the Alfv\'en units on the length scale of $2\pi$.
The magnetic field is derived from the Gauss potential $Q$ of order unity,
and the total energy is of order unity ($\sim 0.7$). 
If we renormalize the length scale to unity, the Alfv\'en velocity is multiplied by $2\pi$.
Therefore, the unit time amounts $4\pi$ Alf\'en time.

\begin{figure}[tb]
\begin{center}
\begin{minipage}{0.5\hsize}
\includegraphics[scale=0.25]{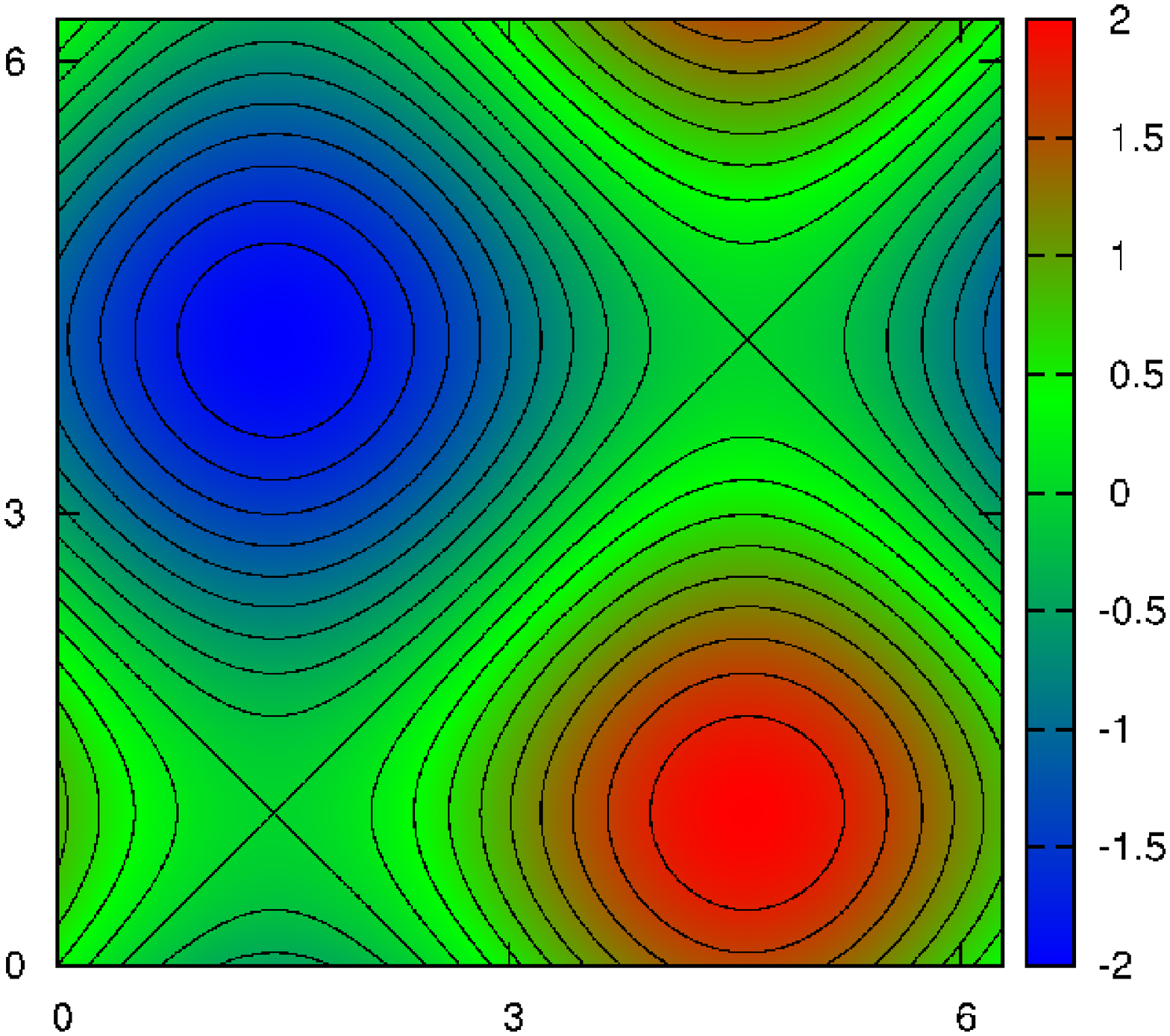}
\begin{center}
\textbf{(a) $t=0.0$}
\end{center}
\includegraphics[scale=0.25]{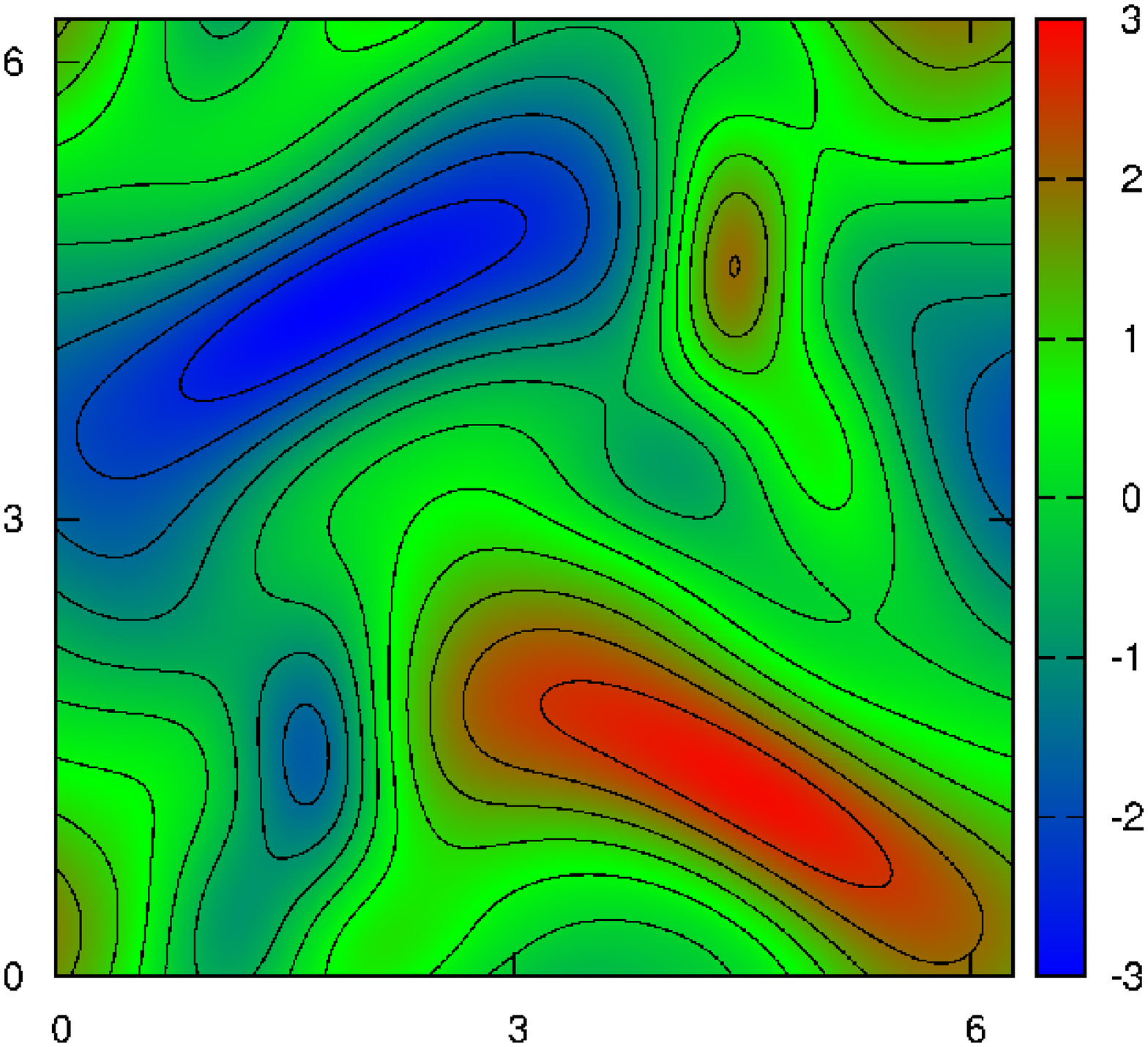}
\begin{center}
\textbf{(b) $t=0.6$}
\end{center}
\end{minipage}
\begin{minipage}{0.5\hsize}
\includegraphics[scale=0.25]{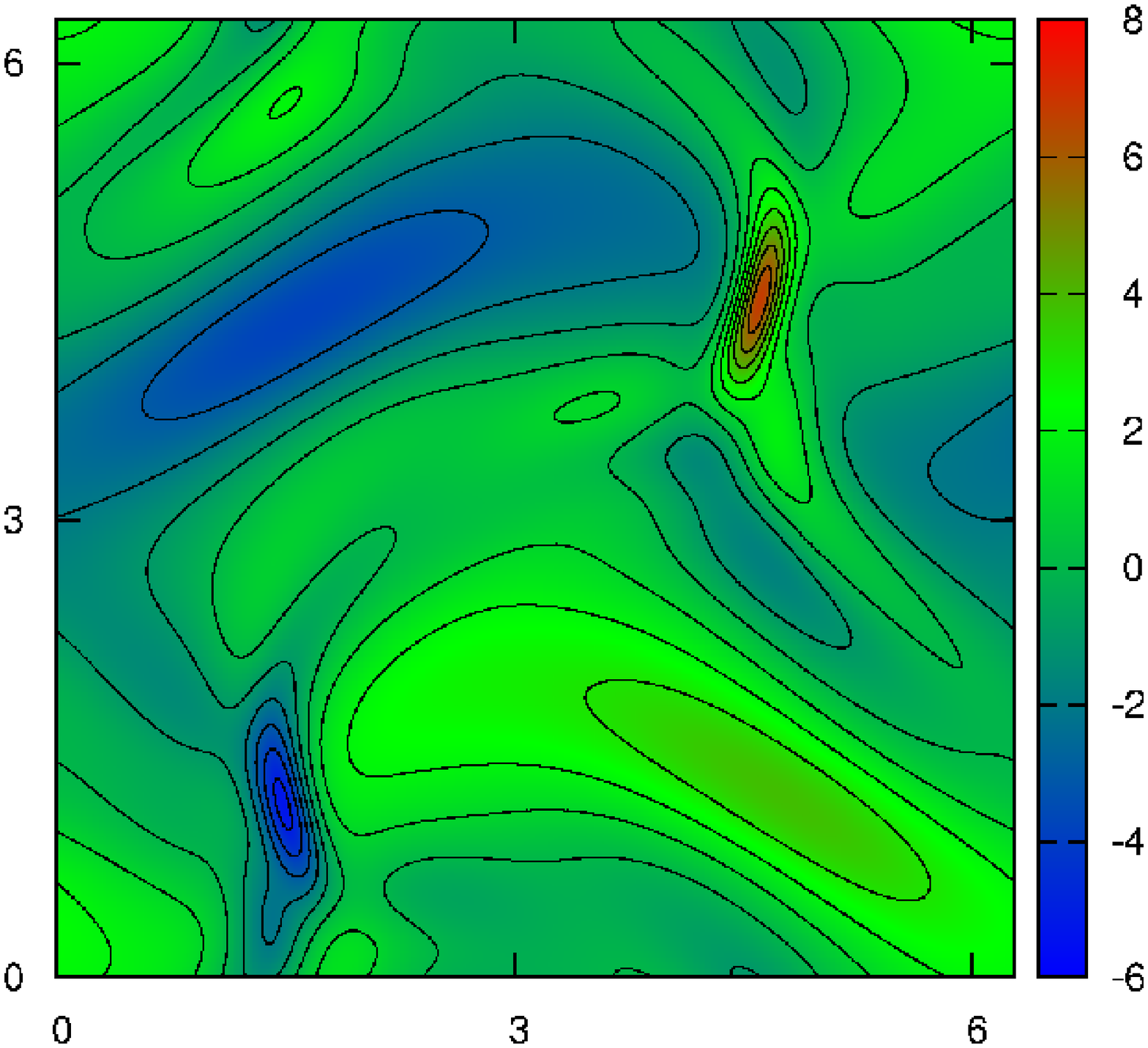}
\begin{center}
\textbf{(c) $t=1.2$}
\end{center}
\includegraphics[scale=0.25]{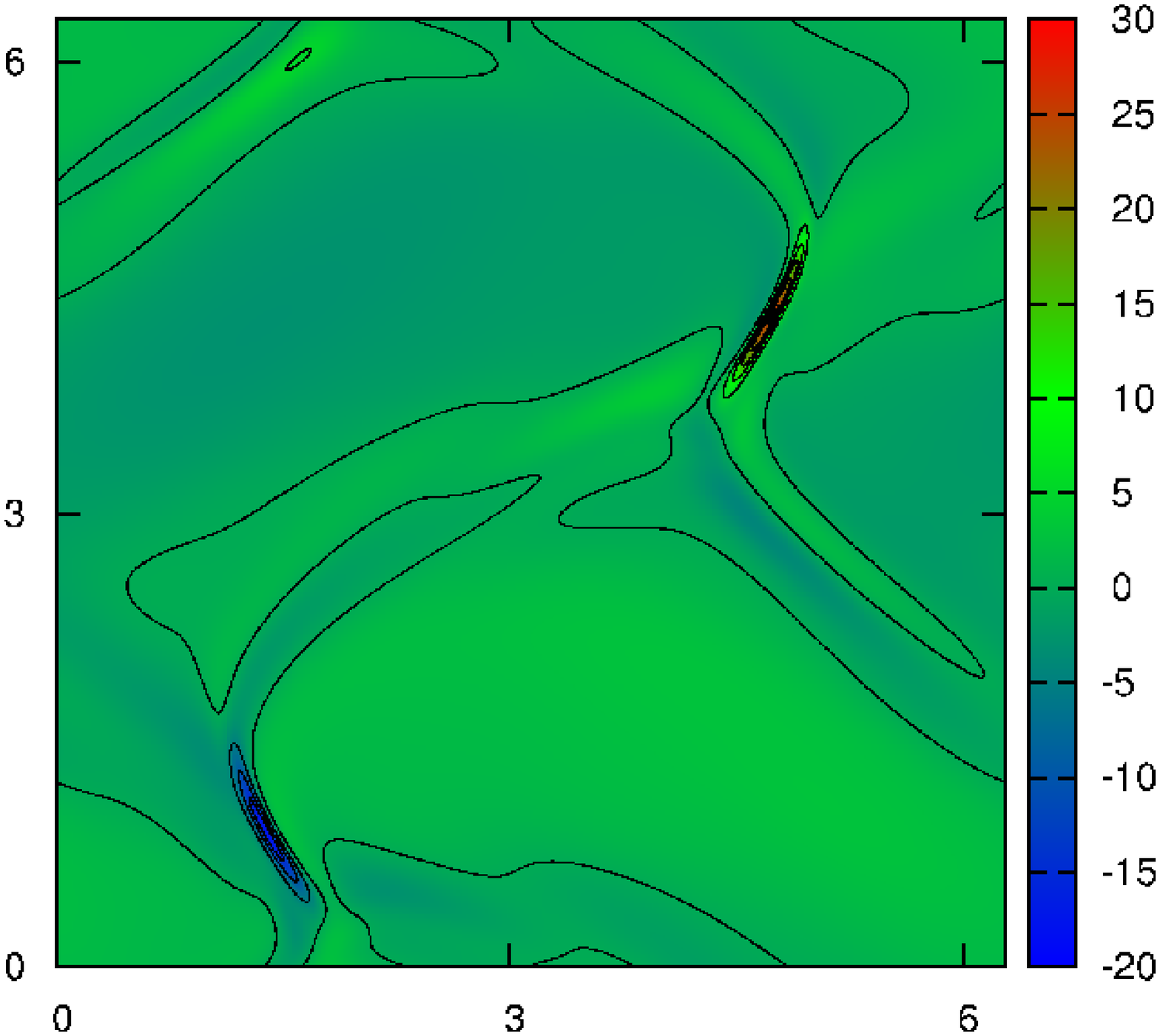}
\begin{center}
\textbf{(d) $t=1.8$}
\end{center}
\end{minipage}
\caption{
\label{evolution} 
Creation of current sheets. The figures show the contours of $J=-\Delta Q$ at different times.
}
\end{center}
\end{figure}

\begin{figure}[tb]
\begin{center}
\includegraphics[scale=0.3]{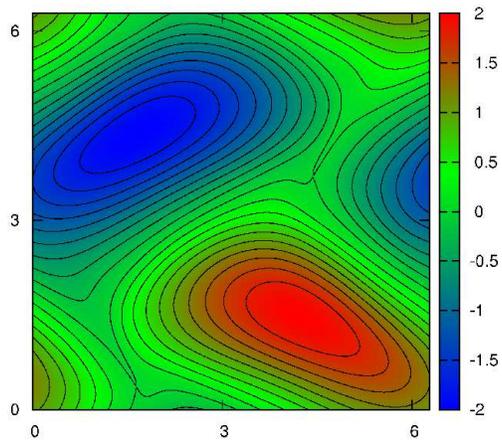}
\caption{
\label{psi}
The contours of the magnetic flux function $\psi=Q$ corresponding to Fig.~\ref{evolution}~(d).}
\end{center}
\end{figure}

Figure\,\ref{evolution} shows the evolution of $J=-\Delta Q$ in RMHD.
We observe that current sheets emerge and stay at almost fixed places, each of which is hemmed between two regions of anti-parallel magnetic fields.
The contours of the magnetic flux function $\psi=Q$ is shown in Fig.\,\ref{psi}.

\begin{figure}[tb]
\begin{center}
\includegraphics[scale=0.3]{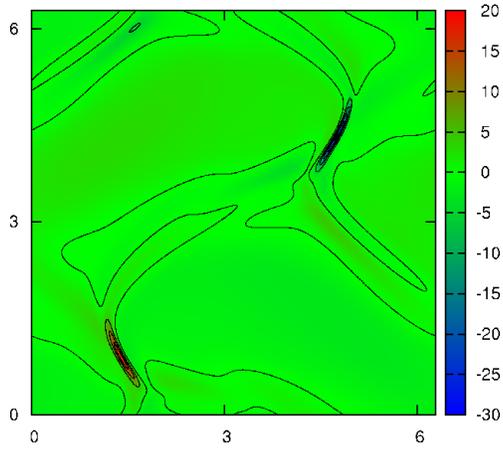}
\begin{center}
(a) RMHD
\end{center}
\includegraphics[scale=0.3]{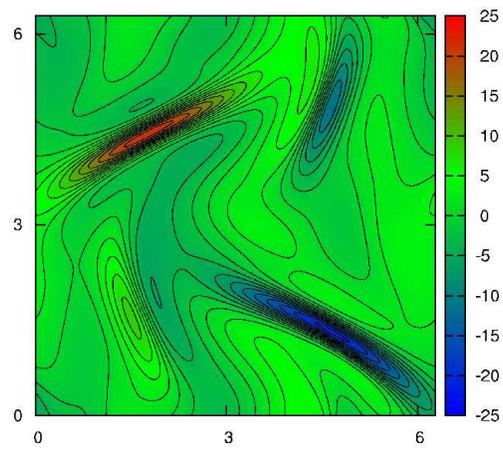}
\end{center}
\begin{center}
(b) 2DEV
\end{center}
\caption{
\label{figJ}
The stills at $t=1.8$ of $\Delta Q$ in the solutions of (a) RMHD and (b) 2DEV. 
We observe a narrower current sheet in RMHD.}
\end{figure}

In Fig.\,\ref{figJ}, we compare $\Delta Q$ in (a) the solution of RMHD, and (b) the solution of 2DEV.
Both figures are the stills at $t=1.8$ of the solutions starting from the same initial condition (\ref{IC}).
As in Fig.\,\ref{evolution}, $-\Delta Q$ of RMHD is the current density $J$.
In 2DEV, however, it does not have a direct physical meaning.
We observe a clear difference between these two systems;
evidently, RMHD produces narrower and stronger current sheets.

\begin{figure}[tb]
\begin{center}
\includegraphics[scale=0.3]{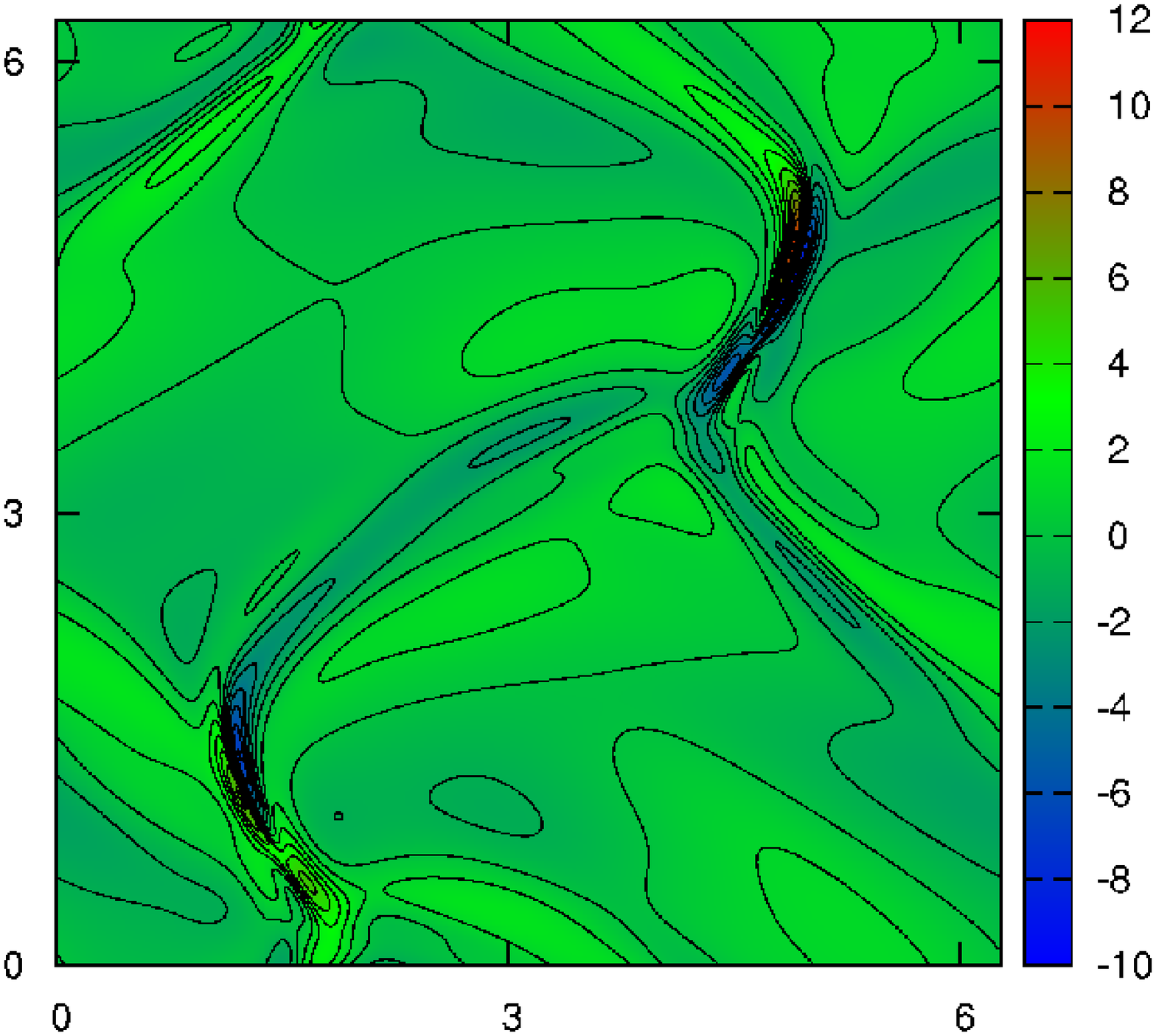}
\begin{center}
(a) RMHD
\end{center}
\includegraphics[scale=0.3]{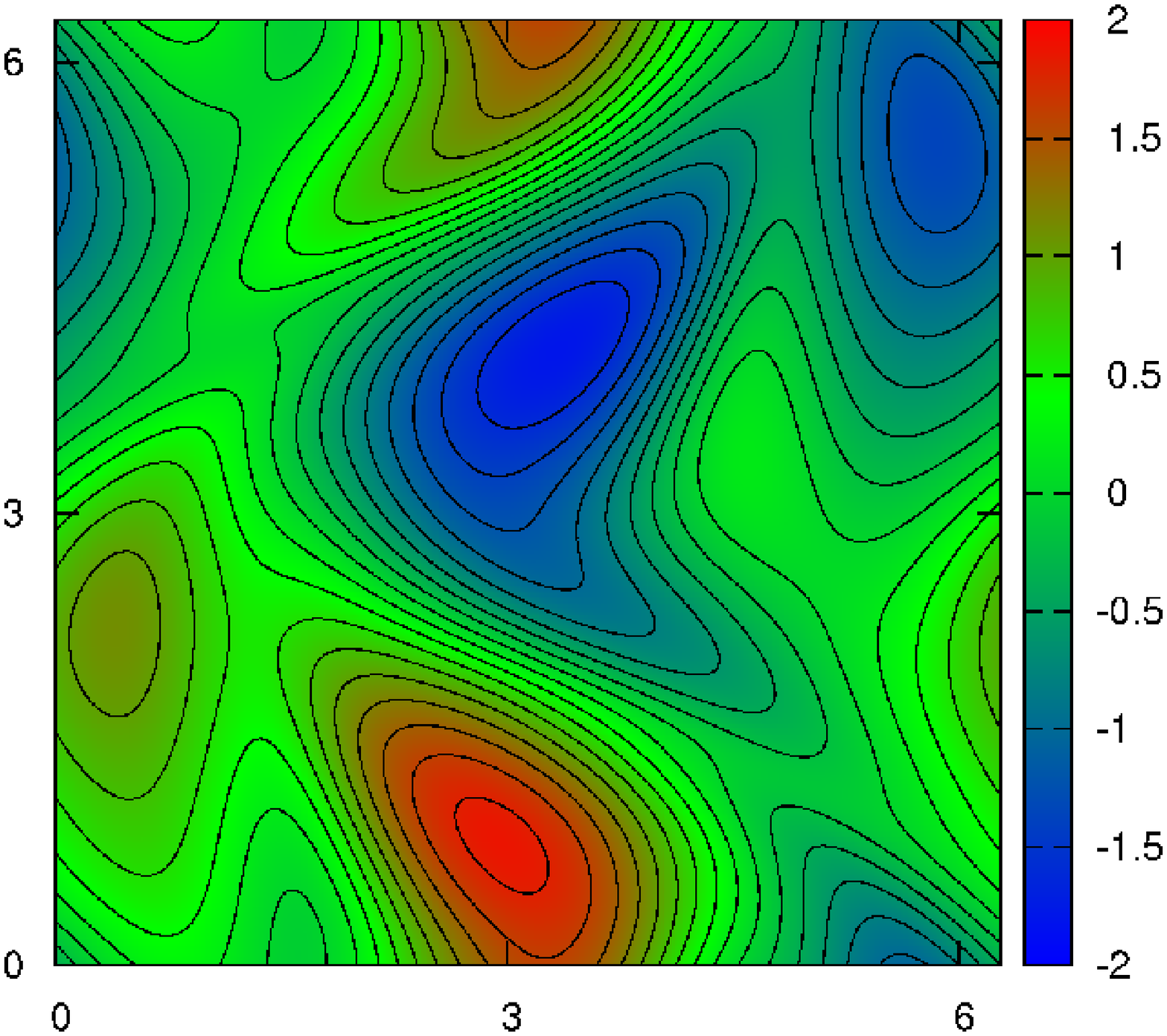}
\begin{center}
(b) 2DEV
\end{center}
\end{center}
\caption{
\label{U}
 The stills at $t=1.8$ of $\omega=[P,Q]$ in the solutions of (a) RMHD and (b) 2DEV.}
\end{figure}

Figure\,\ref{U} (a) shows $\omega=[P,Q]$ of RMHD at $t=1.8$.  
In comparison with $\omega$ of 2DEV, shown in Fig.\,\ref{U} (b), 
we find that stronger vortexes are created in RMHD.
Comparing with Fig.\,\ref{figJ}, we find that the current sheets are created in the
places of localized strong vortexes.

\begin{figure}[tb]
\begin{center}
\includegraphics[scale=0.3]{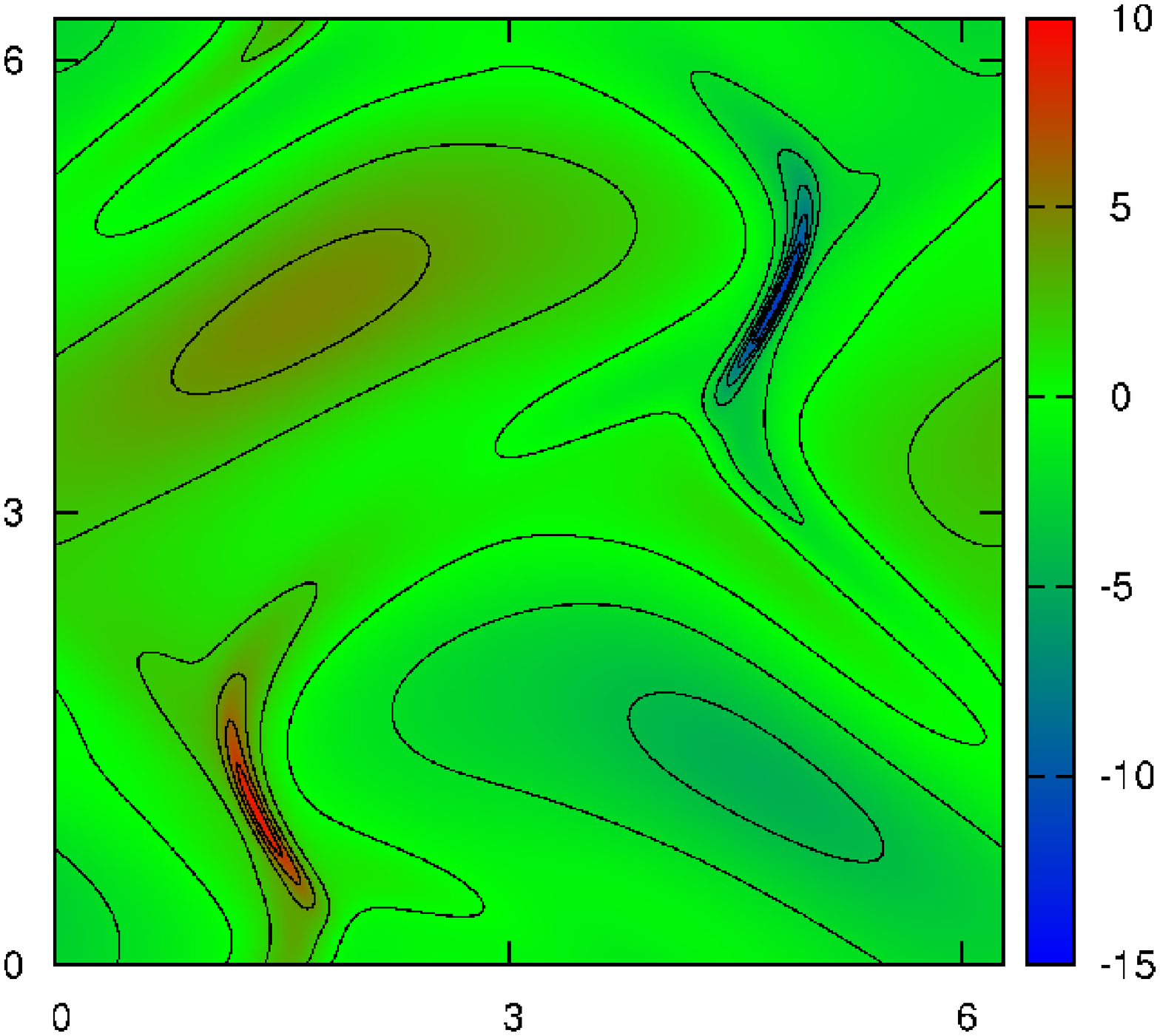} 
\begin{center}
(a) RMHD
\end{center}
\includegraphics[scale=0.3]{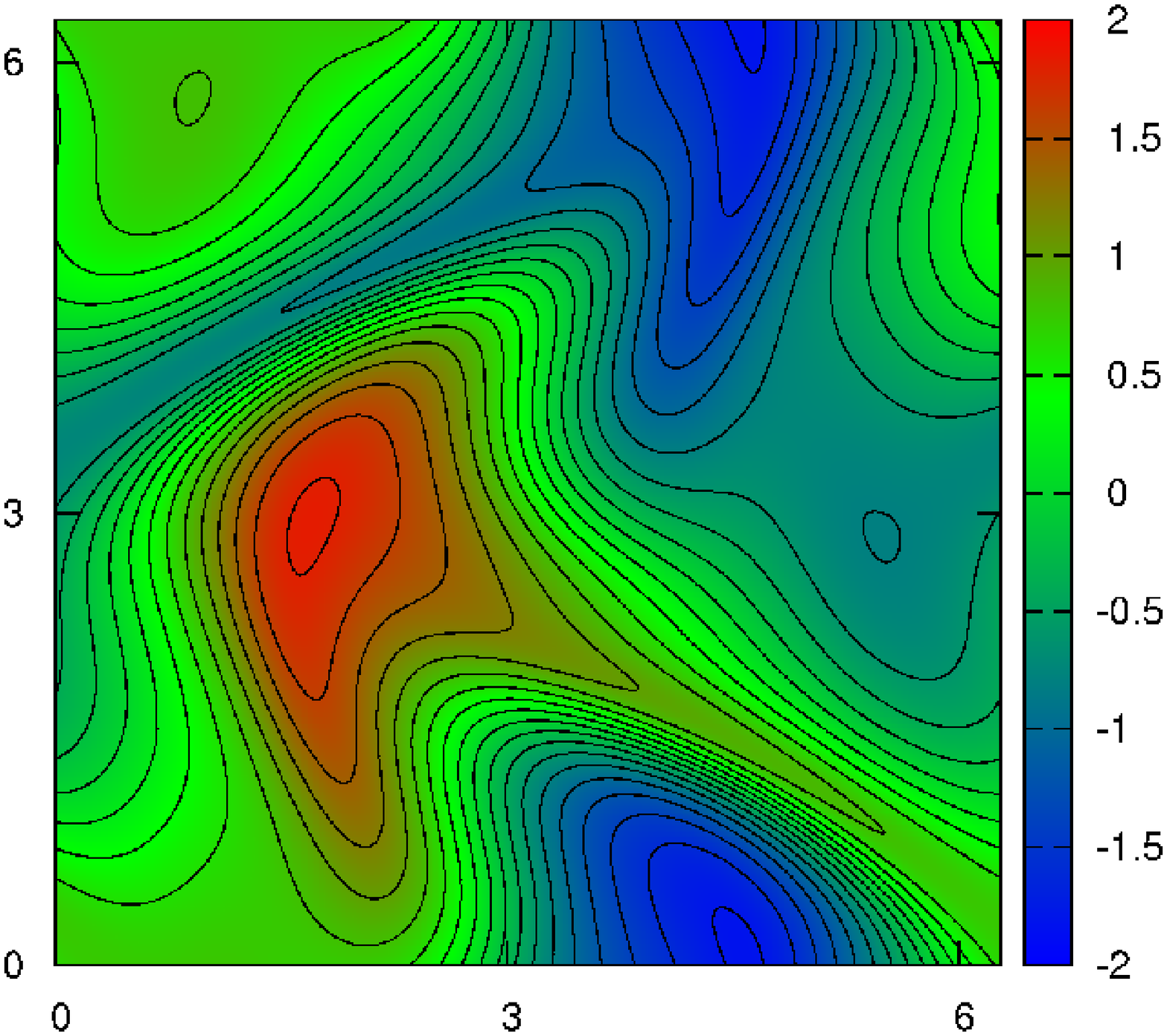}
\begin{center}
(b) 2DEV
\end{center}
\end{center}
\caption{
\label{labelP}
 The stills at $t=1.8$ of $P$ in the solutions of (a) RMHD and (b) 2DEV.}
\end{figure}

Let us close look at the role of the magnetic force
(which is represented by the term $J=-\Delta Q$ in the determining equation (\ref{703-1-P}) of $P$)
in the evolution of the vorticity $\omega$, 
As we discussed in Sec.\,\ref{sec:scaling}, $\int Jdt$ increases $P$ to amplify $\omega=[P,Q]$. 
Figure\,\ref{labelP} shows the stills at $t=1.8$ of $P$ in (a) RMHD and (b) 2DEV,
proving that $P$ peaks at the current sheet in the RMHD solution (see Fig.\,\ref{figJ}).
In the 2DEV solution, we do not find such peaks.

\begin{figure}[tb]
\begin{center}
\includegraphics[scale=0.6, angle=-90]{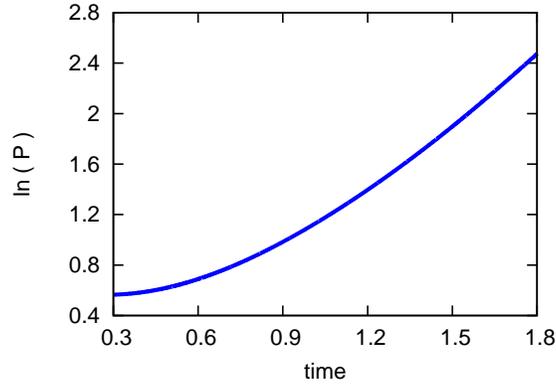}
\begin{center}
(a) $\ln |P|$
\end{center}
\includegraphics[scale=0.6, angle=-90]{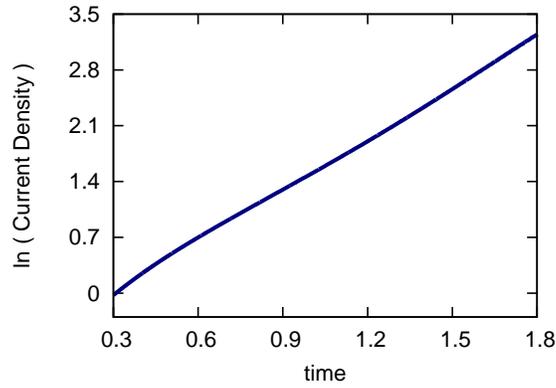}
\begin{center}
(b) $\ln |J|$
\end{center}
\caption{
\label{maxP}
 The evolution of the maximum value of (a) $\ln |P|$ and (b) $\ln |J|$.}
\end{center}
\end{figure}

Now we study the growth rates of $J=-\Delta Q$ and $P$ in the current sheets. 
Figure\,\ref{maxP}\,(a) shows the evolution of the maximum value of $\ln |P|$ in the RMHD system,
which gives a numerical-experimental proof for the heuristic estimate (\ref{P-estimate}).
The asymptotic time constant ($\gamma=2v'_0$) of the exponential growth is about 1.82.
In response to the growth of $|P|$, the current $J$ also glows exponentially;
see Fig.\,\ref{maxP}\,(b).

\begin{figure}[tb]
\begin{center}
\includegraphics[scale=0.5]{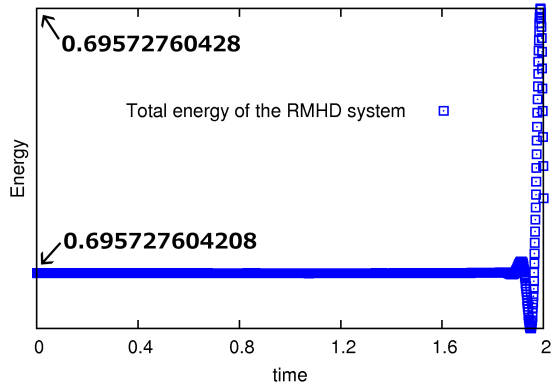}
\end{center}
\caption{
\label{energy} 
The evolution of the total energy of the canonized RMHD system.}
\end{figure}

\begin{figure}[tb]
\begin{center}
\includegraphics[scale=0.425, angle=-90]{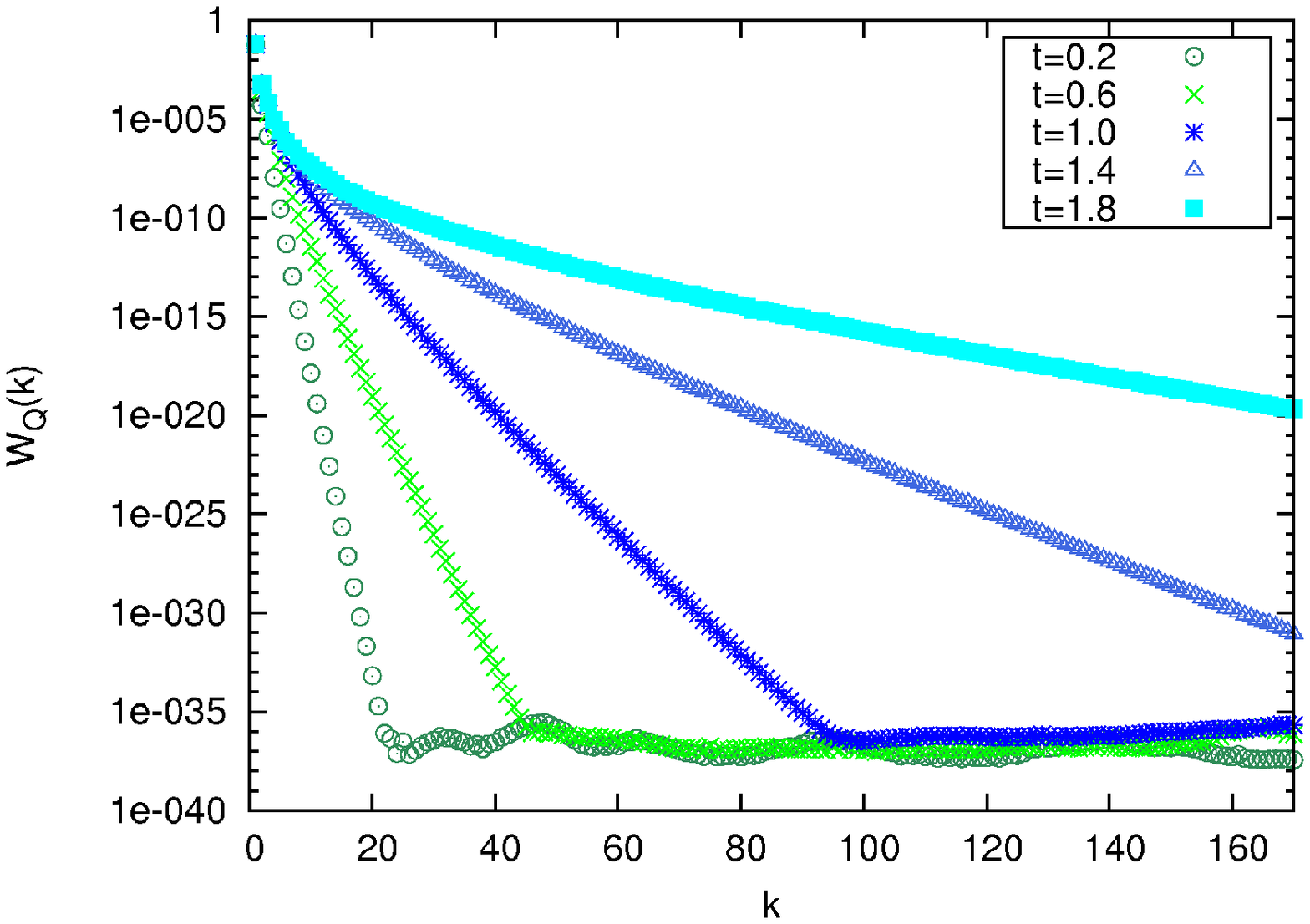}
\begin{center}
(a) RMHD 
\end{center}
\includegraphics[scale=0.425, angle=-90]{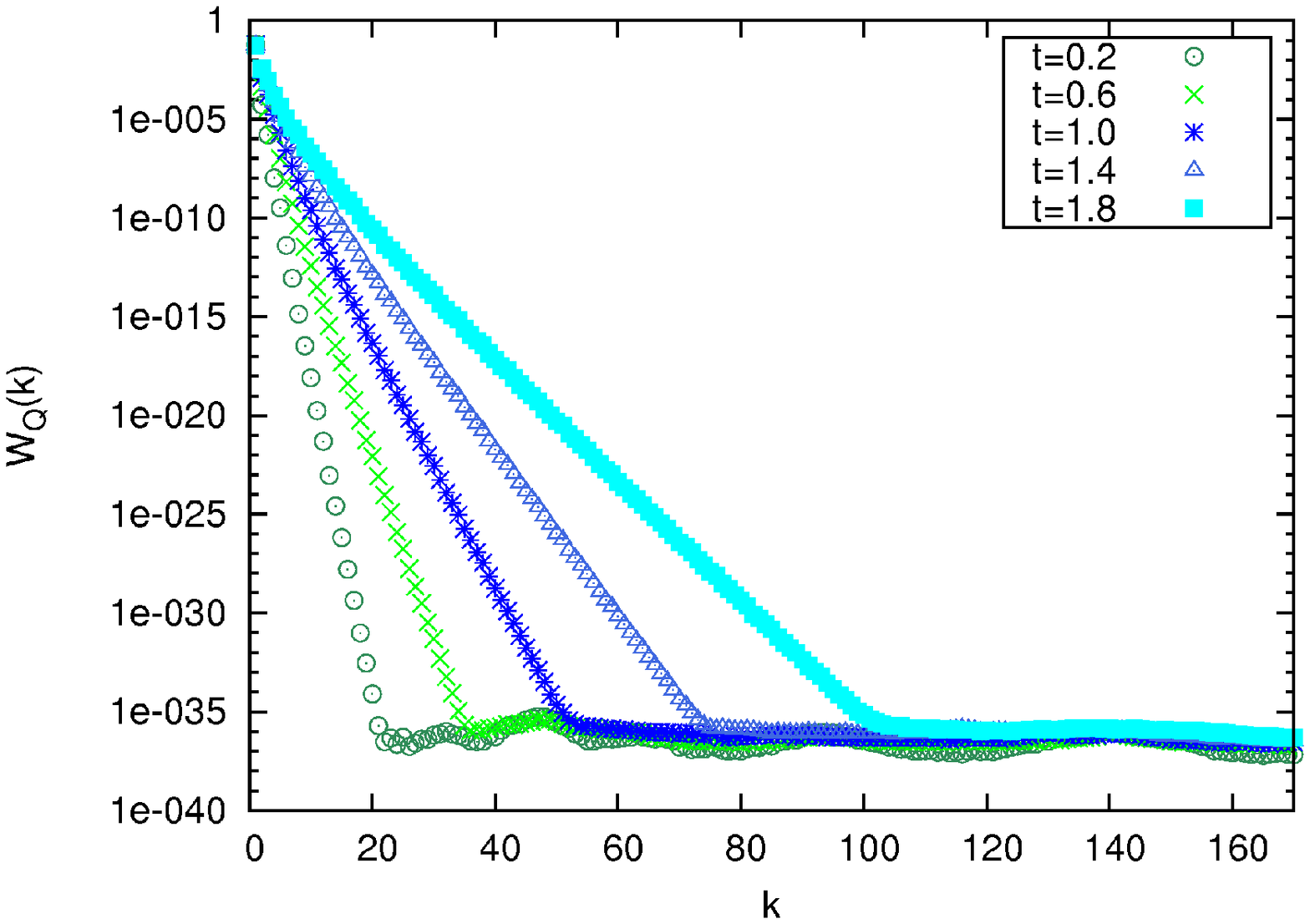}
\begin{center}
(b) 2DEV
\end{center}
\end{center}
\caption{
\label{figsp1} 
The evolution of the wave-number spectra of $Q$ of the solutions of (a) RMHD and (b) 2DEV.}
\end{figure}

\begin{figure}[tb]
\begin{center}
\includegraphics[scale=0.425, angle=-90]{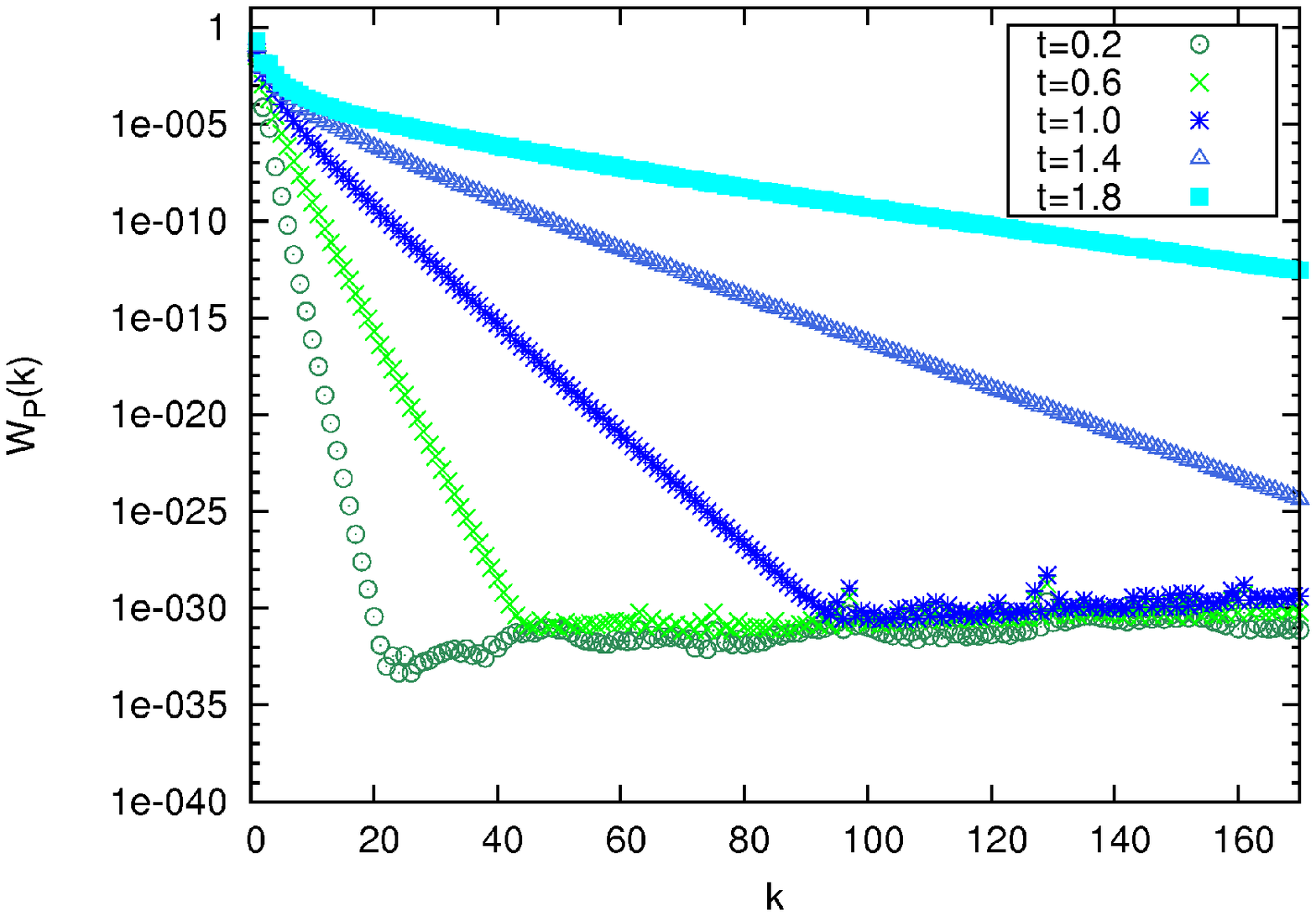}
\begin{center}
(a) RMHD
\end{center}
\includegraphics[scale=0.425, angle=-90]{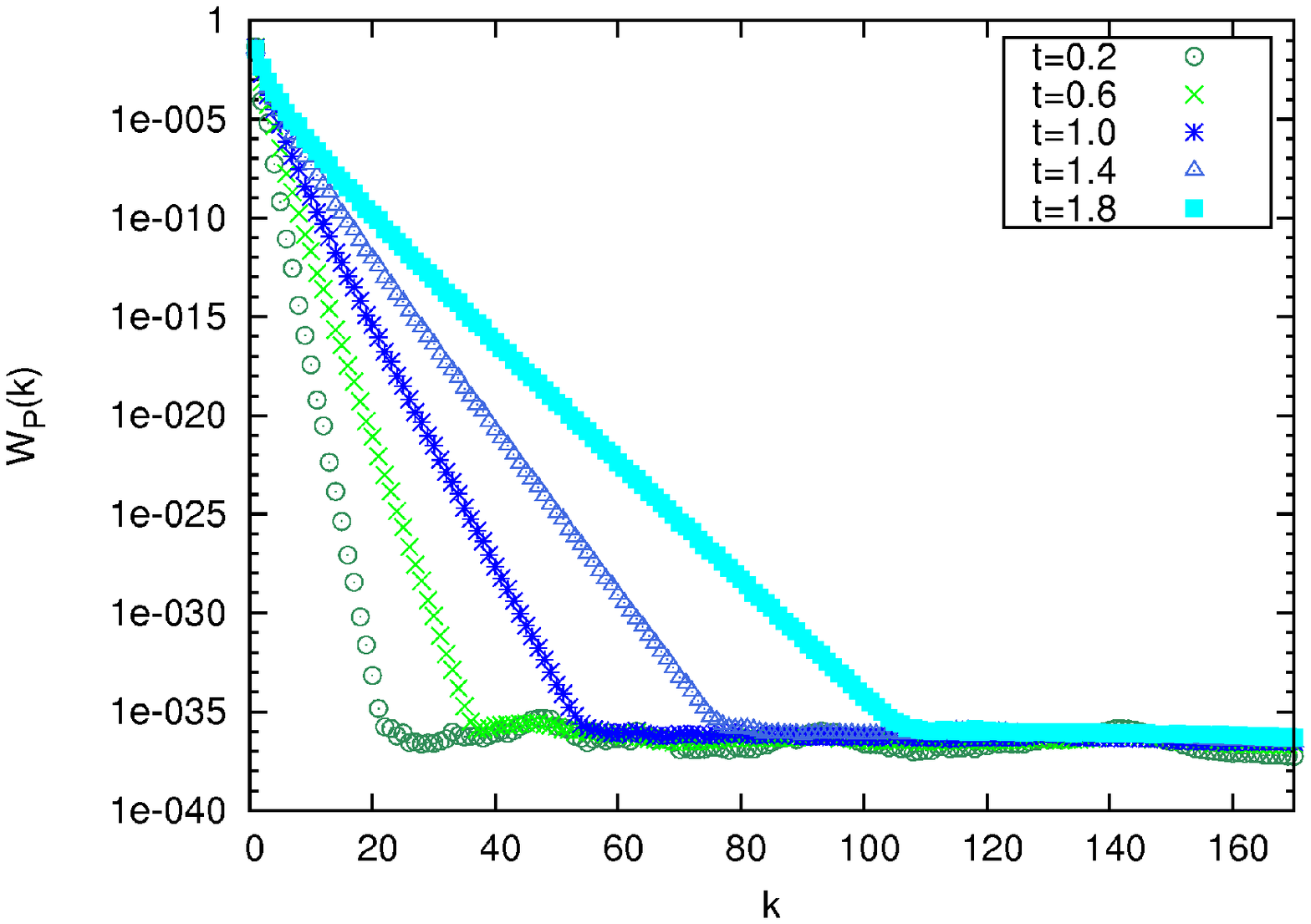}
\begin{center}
(b) 2DEV
\end{center}
\end{center}
\caption{
\label{figsp2} 
The evolution of the wave-number spectra of $P$ of the solutions of (a) RMHD and (b) 2DEV.}
\end{figure}

At the end of this section, we confirm the accuracy of the numerical simulation.
Figure\,\ref{energy} shows the conservation of energy. 
The error is smaller than $10^{-12}$ until $t=1.8$.
Figures\,\ref{figsp1} and \ref{figsp2} respectively show the wave-number spectra of $Q$ and $P$ for the
solutions of (a) RMHD and (b) 2DEV. 
In both variables, the spectra of the RMHD solution become broader with time, reflecting the significant reduction of the length scale due to the creation of localized vorticity $\omega$.

\section{Conclusion}
\label{sec:conclusion}
We have formulated a simple canonized Hamiltonian system 
(\ref{703-1}) that dictates a class of symplectic leaves of RMHD
(the cross helicity is restricted to be zero).
The magnetic force term (which is a nonlinear term of magnetic field in the conventional noncanonical formulation)
is represented by a linear term (just the current $J=-\Delta Q$);
omitting it reduces RMHD into 2DEV.
The vorticity, in turn, is represented by a bilinear form $\omega=[P,Q]$,
which, having the dimension of reciprocal time, determines the rate of change of $J$.
On the other hand, $\int Jdt$ increases $P$ to enhance $\omega$.   
In Sec.\,\ref{sec:scaling}, we have shown that an exponential function $e^{\gamma t}$ 
with a \emph{renormalized} time constant $\gamma$ can describe a self-consistent evolution of $\Delta Q$ and $P$.
By numerical simulation, we have demonstrated that exponentially growing $J$ creates current sheets in the RMHD system.
In the 2DEV system, in which the term $\int Jdt$ is absent in $P$, $\Delta Q$ glows slower than the RMHD system.

\begin{figure}[tb]
\begin{center}
\includegraphics[scale=0.5]{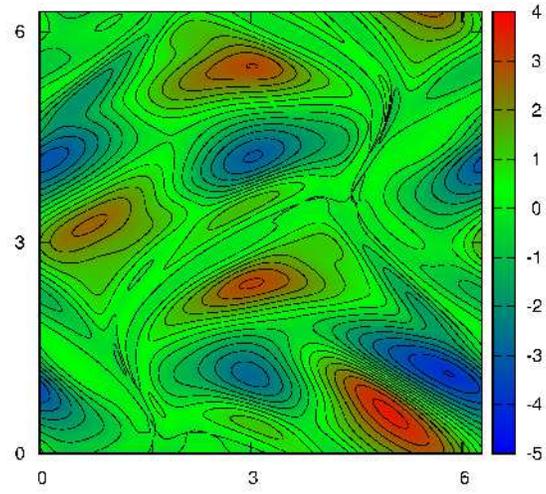} 
\end{center}
\caption{
\label{figCD}
 The stills at $t=1.8$ of $\omega g(\psi)=[P,Q^2]$ in the solution of RMHD.}
\end{figure}

We end this paper with some comments on the limitations of this work.
As shown in (\ref{cross-helicity-QP-form}), 
our model describes the dynamics on the zero cross helicity submanifolds of the phase space.
We note that this topological constraint does not mean that the local value of 
$\omega g(\psi)= [P,G(Q)]$ is zero; see Fig.\,\ref{figCD},
only the integral over the total domain must vanish.
By the completeness of the Clebsch parameterization (\ref{c01-1}) of the closed 2-form $\omega$\,\cite{refZY09}
(the specialty of (\ref{c01-1}) yielding zero cross helicity is in that one of the Clebsch parameter is chosen to be $Q=\psi$, but the other parameter $P$ may cover the entire degree of freedom of the single variable $\omega$),
we may expect that the local dynamics near each current sheet is not influenced by the zero cross helicity constraint.  
However, a definite conclusion awaits careful comparisons of zero and finite cross helicity cases. 
We may formulate the latter by invoking an extended set of canonical variables; see, for example, the formulation by Morrison and Hazeltine~\cite{refPJ84}.
Detailed results will be discussed elsewhere.
We also note that Hamiltonian formalisms do not include the effect of resistivity and viscosity.
Some authors (for example~\cite{Kishimoto,Kishimoto2}) report super-exponential (like $\exp( a e^{bt})$) evolution in the event of magnetic reconnection.
Similar fast growths have been observed in a system of finite electron inertia~\cite{refOP}.
However, it seems that such growth is not a long term behavior; 
instead, the super-exponentiality is due to a slower growth in the initial phase, because the vorticity must
develop, from a very small perturbation, by a resistive instability.
It is not simple to include a resistivity term in the present formulation using the $Q$-$P$ canonical variables.
However, the finite electron inertia model has a Hamiltonian structure~\cite{refPe,refPe2}, 
thus it might be the target for extending this work to the study of ``collision-less'' reconnection.

\section*{Acknowledgment}
This work was supported by a Grant-in-Aid for Scientific Research Japanese Ministry of Education, Science and Culture 
(23224014) from MEXT, Japan.

\clearpage

\end{document}